\documentclass[aps,preprint,prd,showpacs,nofootinbib]{revtex4}

\usepackage{latexsym}
\usepackage{amsmath}
\usepackage{graphicx}
\usepackage{subfigure}
\usepackage{dcolumn}
\usepackage{bm}
\usepackage{amssymb}
\usepackage{latexsym}
\usepackage{ulem}
\usepackage{color}
\usepackage[colorlinks,linkcolor=magenta,anchorcolor=cyan,citecolor=blue]{hyperref}

\def\be{\begin{equation}}
\def\ee{\end{equation}}
\def\ba{\begin{eqnarray}}
\def\ea{\end{eqnarray}}

\def\blue{\color{blue}}

\def\lf{\left}
\def\rt{\right}

\begin{document}

\title{ Is the NANOGrav signal a hint of dS decay during inflation?}

\author{Hao-Hao Li$^{1}$\footnote{\texttt{\blue lihaohao18@mails.ucas.ac.cn}}}
\author{Gen Ye$^{1}$\footnote{\texttt{\blue yegen14@mails.ucas.ac.cn}}}
\author{Yun-Song Piao$^{1,2,3,4}$\footnote{\texttt{\blue yspiao@ucas.ac.cn}}}

\affiliation{$^1$ School of Physics, University of Chinese Academy of
Sciences, Beijing 100049, China}

\affiliation{$^3$ Institute of Theoretical Physics, Chinese
Academy of Sciences, P.O. Box 2735, Beijing 100190, China}

\affiliation{$^3$ School of Fundamental Physics and Mathematical
Sciences, Hangzhou Institute for Advanced Study, UCAS, Hangzhou
310024, China}

\affiliation{$^4$ International Center for Theoretical Physics
Asia-Pacific, Beijing/Hangzhou, China}

\begin{abstract}

The NANOGrav collaboration recently has reported evidence for
a stochastic common-spectrum process in its 12.5-yr dataset. We
show that such a signal might be a hint of de Sitter (dS) decay during
inflation. As suggested by the swampland conjectures, dS space is
highly unstable if it exists at all. During inflation, the
short-lived dS states will decay through a cascade of
first-order phase transitions (PT). We find that the gravitational
waves (GWs) yielded by such a PT will be ``reddened" by subsequent
dS expansion, which may result in a slightly red-tilt stochastic
GW background at the low-frequency band, compatible with the NANOGrav
12.5-yr result.

\end{abstract}

\maketitle



The primordial GW background (GWB)
\cite{Starobinsky:1979ty,Rubakov:1982df} spans a broad
frequency band, $10^{-18}-10^{10}$Hz,
e.g.\cite{Lasky:2015lej,Wang:2016tbj,Caprini:2018mtu}. It is
usually thought that the discovery of primordial GWs will solidify
our confidence on inflation. The primordial GWs at ultra-low
frequency $10^{-18}-10^{-16}$Hz may induce B-mode polarization
in the cosmic microwave background (CMB)
\cite{Kamionkowski:1996ks,Kamionkowski:1996zd}. The search for the
primordial GWs with CMB has been still on the way.

The Pulsar Timing Array (PTA) experiments, sensitive to GWs with
frequency $f\sim 1/yr$, have also been searching for such a
background. Recently, the NANOGrav collaboration
\cite{Arzoumanian:2020vkk}, based on the analysis of $12.5$-yr
data, reported evidence for a stochastic common-spectrum process
with frequency $f\sim 1/yr$ ($\sim 10^{-8}$Hz), which might be
interpreted as a stochastic GWB with a spectrum $\Omega_{GW}\sim
f^{-1.5\sim 0.5}$ at $1\sigma$ level.
It is actually difficult for inflation to yield such a stochastic
GWB, see recent \cite{Vagnozzi:2020gtf}, which seems to require a
highly blue-tilt GW spectrum e.g.\cite{Piao:2004tq,Baldi:2005gk},
see also \cite{Cai:2016ldn,Cai:2015yza}.


It is well-known that evolution of the Universe must be
described in an effective field theory (EFT) that has a UV-completion.
Recently, the Trans-Planckian Censorship Conjecture (TCC) has been
proposed in Ref.\cite{Bedroya:2019snp}, which states that the
sub-Planckian fluctuations will never have its length scale larger
than the Hubble scale, otherwise the EFTs will belong to the swampland
(without UV-completion).
The swampland conjectures \cite{Ooguri:2006in,Obied:2018sgi}
actually suggest that dS space is highly unstable if it exists at
all, while AdS phase is ubiquitous, see
e.g.\cite{Ye:2020btb,Ye:2020oix} for its implications to the
observable Universe. According to TCC,  inflation in the early
Universe can only last for a limited e-folding number $\int H
dt<\ln{M_{{P}}\over H}$ \cite{Bedroya:2019tba}, see also
Refs.\cite{Li:2019ipk,Berera:2019zdd,Dhuria:2019oyf,Torabian:2019zms}
for multi-stage inflation. It has been proposed in
Refs.\cite{Bedroya:2019snp,Bedroya:2020rac} that, during inflation, the short-lived
dS state($\Delta t< \frac{1}{H} \ln{M_{{P}}\over H}$)  will decay to a dS state
with lower energy through the non-perturbative nucleation of
bubbles \cite{Coleman:1980aw}, so a cascade of dS decay will be
present, see also earlier
\cite{Freese:2004vs,Freese:2005kt,Liu:2009pk}. It is significant
to ask if such short-lived dS vacua have any observable imprint.

We will present this possibility. A novelty of our result is that
the NANOGrav signal, if being a stochastic GWB, will be a hint of
dS decay during inflation. The first-order PT that dS bubbles nucleate and collide
yields a sub-horizon GWB with a peculiar spectrum $P_T^{PT}$
\cite{Kosowsky:1991ua,Caprini:2007xq,Huber:2008hg}, it will alter
the initial state of GW mode if $P_T^{PT}\gg P_T^{BD}$ for ${k\gg
aH}$, where $P_T^{BD}$ is the Bunch-Davis spectrum. Usually, the
low-frequency GWB (at PTA band) requires low-energy PT.
Here, though the PT occurred during inflation, which is at a
high-energy scale, the subsequent inflation will stretch the
corresponding sub-horizon GW mode outside the horizon, which not
only redshifts their frequency but also reddens their spectra
\cite{Wang:2018caj,Jiang:2015qor}. We will show how such a
scenario works.

\begin{figure}[htbp]
\includegraphics[scale=2,width=0.7\textwidth]{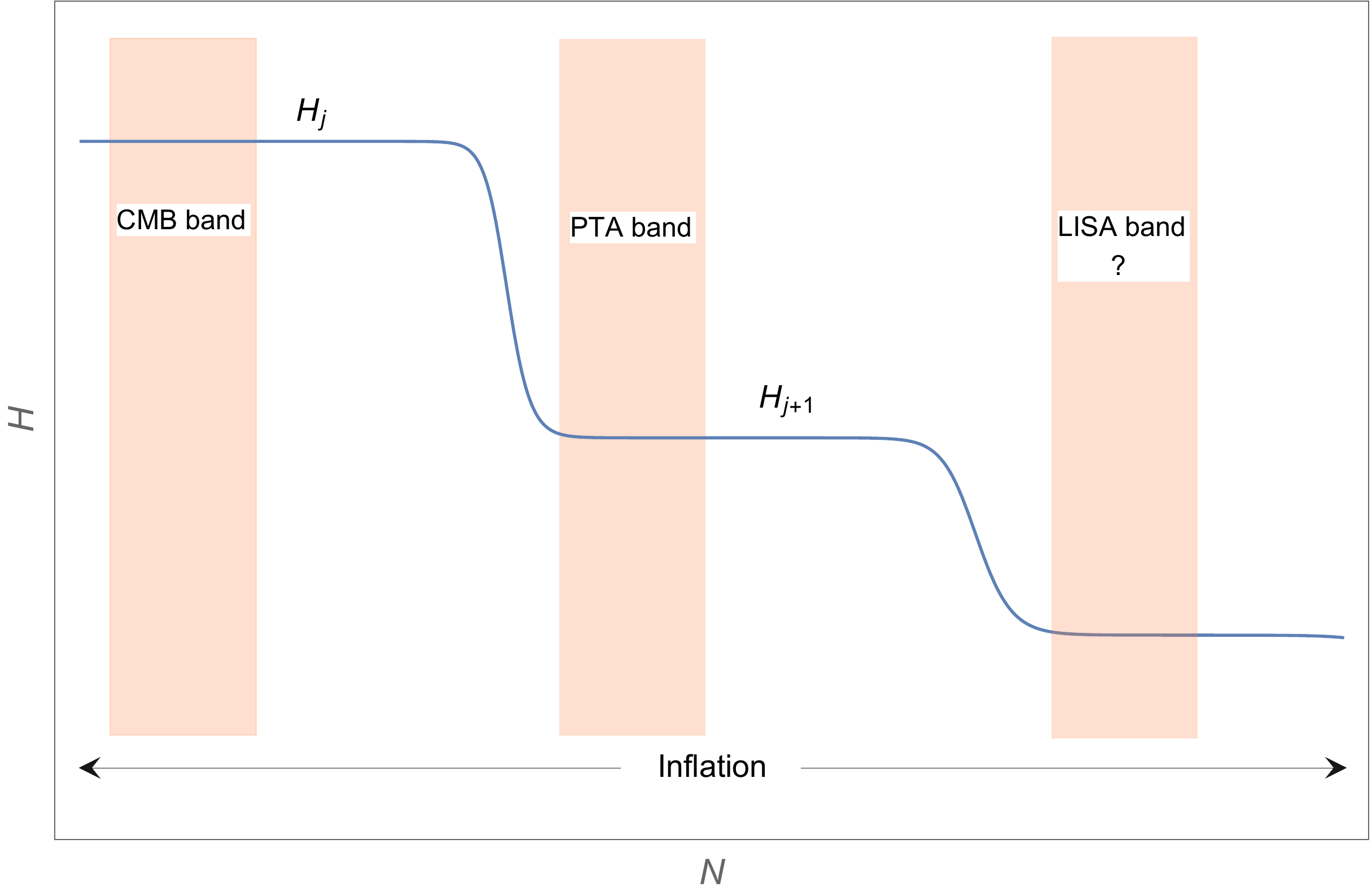}
\caption{A sketch of the dS cascade. Initially, the inflation with
$\rho=\Lambda_{j}$ occurred. After the first-order PT, inflation continues but with lower $\rho=\Lambda_{j+1}<\Lambda_{j}$,
and so on. }
\label{fig-PT}
\end{figure}

The scenario we consider is sketched in Fig-\ref{fig-PT}. The
e-folding number that the $j$-th stage of
inflation lasts is bounded by $N_j\simeq H_j\Delta
t_j<\ln{M_{{P}}\over H_j}$, where $H_j^2={\Lambda_j\over 3M_P^2}$
with $\Lambda_{j}$ being the vacuum energy of the
$j$-th stage. During $\Delta t_j$, a first-order PT
must occur. Until ${\Gamma_j/ H^4_j}\gtrsim 1$
\cite{Turner:1992tz}, the PT completes, where $\Gamma_j\sim
e^{\beta(t-t_*)}$ is the nucleating rate of bubbles with
vacuum energy $\Lambda_{j+1}$. The relevant physics underlying the PT
is encoded in $\beta$, which is not our focus here (as an example, we present a phenomenological model in the Appendix). When the bubbles
collide, the energy of bubble walls is efficiently released,
e.g.\cite{Watkins:1991zt,Kolb:1996jr,Zhang:2010qg}, and rapidly
diluted with the expansion of the Universe. Hereafter, the
$(j+1)$-th stage inflation with $\rho=\Lambda_{j+1}$
will start, and last $N_{j+1}<\ln{M_{{P}}\over H_{j+1}}$ until
${\Gamma_{j+1}/ H^4_{j+1}}\gtrsim 1$.

According to the swampland conjectures, just one-stage slow-roll
inflation could not bring a large enough e-folding number $ N $,
so the multi-stage inflation model in Fig-\ref{fig-PT} is interesting. We
present a phenomenological model of multi-stage inflation with the
first-order PT in the Appendix, in which each stage of inflation
satisfies the swampland conjectures.

The tensor perturbation is $\gamma_{ij}(\tau,\mathbf{x})=\int
\frac{d^3k}{(2\pi)^{3} }e^{-i\mathbf{k}\cdot \mathbf{x}}
\sum_{\lambda=+,\times} \hat{\gamma}_{\lambda}(\tau,\mathbf{k})
\epsilon^{(\lambda)}_{ij}(\mathbf{k})$,  where $
\hat{\gamma}_{\lambda}(\tau,\mathbf{k})=
\gamma_k(\tau)\hat{a}_{\lambda}(\mathbf{k}) +c.c.$. Its equation
of motion is \be
\frac{d^2u_k}{d\tau^2}+\left(k^2-\frac{a^{\prime\prime}}{a}
\right)u_k=0, \label{eom1} \ee where ${u}_k= {aM_p\over
2}\gamma_{k}$. Initially, the GW modes should be deep inside the horizon, i.e., $k^2 \gg \frac{a^{\prime\prime}}{a}$, so the
initial state is $u_k=C(k)e^{-i k\tau}$ with
$C=\frac{1}{\sqrt{2k}}$.

For a stochastic gravitational wave background, it is not
contested that the power law of energy spectrum  on the low
frequency satisfies $ \Omega_{GW}(k) \propto k^3 $ due to
causality\cite{Caprini:2009fx}. It is hard to calculate a concrete
expression of the energy spectrum $ \Omega_{GW}(k) $ for a complex
source if one does not do some special assumptions. Thus for the PT
source of GWs, we follow the numerical simulation results in
Refs.\cite{Huber:2008hg, Weir:2016tov, Cutting:2018tjt}. In our
model the next stage of inflation begins after a PT, so, instead of the Bunch-Davies vacuum, our initial condition is set
by PT
\be
\Omega_{GW}^j(k)=\Omega_{GW,c}{({\cal A}+{\cal B})k^{\cal
B}_{c}k^{\cal A}\over {\cal B}k^{({\cal A}+{\cal B})}_{c}+{\cal
A}k^{({\cal A}+{\cal B})}}\,, \label{Omega-t} \ee \be
\Omega_{GW,c}=\kappa^2 \lf({\Delta \Lambda_{j}\over
\Lambda_{j}}\rt)^2\lf({H_{j}\over \beta}\rt)^2 {0.11v_b^3\over
0.42+v_b^2}, \ee where $\kappa\simeq 1$ for $\Lambda_{j+1}\ll
\Lambda_{j}$, the bubble being sub-horizon requires $H_{j}/\beta<1$,
$v_b$ is the bubble wall velocity and $\Delta
\Lambda_{j}=\Lambda_{j}-\Lambda_{j+1}$.  According to the numerical
simulation, on the low frequency $ k < k_c \approx a_* \beta  $ ,
${\cal A} = 3$  and ${\cal B} = 1$ are precise enough. Thus at the beginning of
the $(j+1)$-th stage inflation, the initial state $u_k$ of GWs modes will
be inevitably modified as (\ref{Omega-t}).

The energy density of GW is \cite{Boyle:2005se} \be
\rho_{GW}=\sum_{\lambda=+,\times}\rho_{GW}^{\lambda}={ M_p^2\over
4}\int {k^3\over
2\pi^2}\lf({|{\gamma}^\prime_k|^2+{k^2}|\gamma_k|^2\over
a^2}\rt)d\ln{k}.\label{rho-GW}\ee According to
$\Omega_{GW}^j={d\rho_{GW}\over \rho_{j} \lf(d\ln{k}\rt)}$, we
have for sub-horizon modes \be |C(k)|^2={3\pi^2 M_p^2 H_{j}^2}{a^4_*\over
k^5}\Omega_{GW}^j,\label{eq:Ck} \ee where $a_*$ is the
scale factor at PT and $\rho_j=\Lambda_j$. The
inflation with $\Lambda_{j+1}<\Lambda_j$ will start after the PT
completes. Thus the sub-horizon GWs (with wavelength
$\lambda<H_{j}^{-1}\ll H_{j+1}^{-1}$) will be stretched outside
the horizon $1/H_{j+1}$. By requiring that the solution of
Eq.(\ref{eom1}) in the sub-horizon limit is $u_k=C(k)e^{-i
k\tau}$, we obtain $u_k({\tau})={-}C(k)\sqrt{-\pi k{\tau}\over
2}H^{(1)}_{3/2}(-k{\tau})$ with $C(k)$ being (\ref{eq:Ck}). On
super-horizon scale, the Hankel function $
H^{(1)}_{3/2}(-k{\tau})\overset{-k{\tau}\rightarrow0}
    \approx-i \sqrt{2/(-\pi k^3\tau^3)}$. Thus the
    primordial GW spectrum is \be P_T=\frac{4k^3}{\pi^2 M_p^2 a^2} \lf|u_k
    \rt|^2= {12 H^2_{j} H^2_{j+1} \over (k/a_*)^4}\Omega_{GW}^j.
    \label{pt} \ee
The physics of short-lived dS space is encoded in $\Omega_{GW}^j$,
so $P_T$. We see that the sub-horizon state (\ref{Omega-t}) is
reddened by the $(j+1)$-stage inflation, $P_{T}\sim
\Omega_{GW}^j/k^4$, so $P_T$ also records the ``reddening"
character of inflation.

We have $v_b=1$ for $\Lambda_{j+1}\ll \Lambda_{j}$, noting that
the swampland conjecture \cite{Bedroya:2020rac} requires
$\Delta\Lambda_j\simeq \Lambda_j$. We set ${\cal A}\simeq 3$ and
${\cal B}\simeq 1$
\cite{Kosowsky:1991ua,Caprini:2007xq,Huber:2008hg} (see also
\cite{Weir:2016tov,Jinno:2016vai}) in (\ref{Omega-t}). In
addition, the peak momentum of $\Omega_{GW}^j$ is ${k_{c}\over
2\pi \beta a_*}= 0.62/(1.8-0.1v_b+v_b^2)$ \cite{Huber:2008hg},
which suggests $k_{c}\simeq 1.4\beta {a_*}$ for $v_b\simeq 1$.
Thus Eq.(\ref{pt}) becomes \be P_T\simeq
\lf({\beta^{-1}H_{j}}\rt)^6\lf({\Lambda_{j+1}\over
\Lambda_{j}}\rt)\cdot {\lf({k\over k_{c}}\rt)^{-1}\over
1+3\lf({k\over k_{c}}\rt)^{4}}, \label{pt1}\ee where $V_{j/j+1}=
3M_P^2H^2_{j/j+1}$. The wavelength of GW mode being initially
sub-horizon suggests a low-frequency cutoff
$k_{cutoff}={a_*}{H_{j}}$ for $P_T$. We have
$k_{c}/k_{cutoff}\simeq 1.4\beta/H_{j}$, which is consistent with the
requirement that the bubble is sub-horizon, $H_{j}/\beta<1$.
According to (\ref{pt1}), $P_T\sim k^{-5}$ for $k\gg k_{c}$ is
strongly red, while $\sim k^{-1}$ for $k\ll k_{c}$. The maximal
value of $P_T$, i.e.,  $P_{T,max}$, is at $k= k_{cutoff}$. We have
\be P_{T,max}\simeq
\lf({\beta^{-1}H_j}\rt)^{5}\lf({\Lambda_{j+1}\over
\Lambda_{j}}\rt). \label{pt2}\ee
Thus if ${\Lambda_{j+1}/ \Lambda_{j}=0.2}$ and ${H_{j}/\beta}\sim
0.4$, we will have $P_{T,max}\sim 10^{-3}$, far larger than that
in slow-roll inflation scenarios.

\begin{figure}[htbp]
\includegraphics[width=0.484\textwidth]{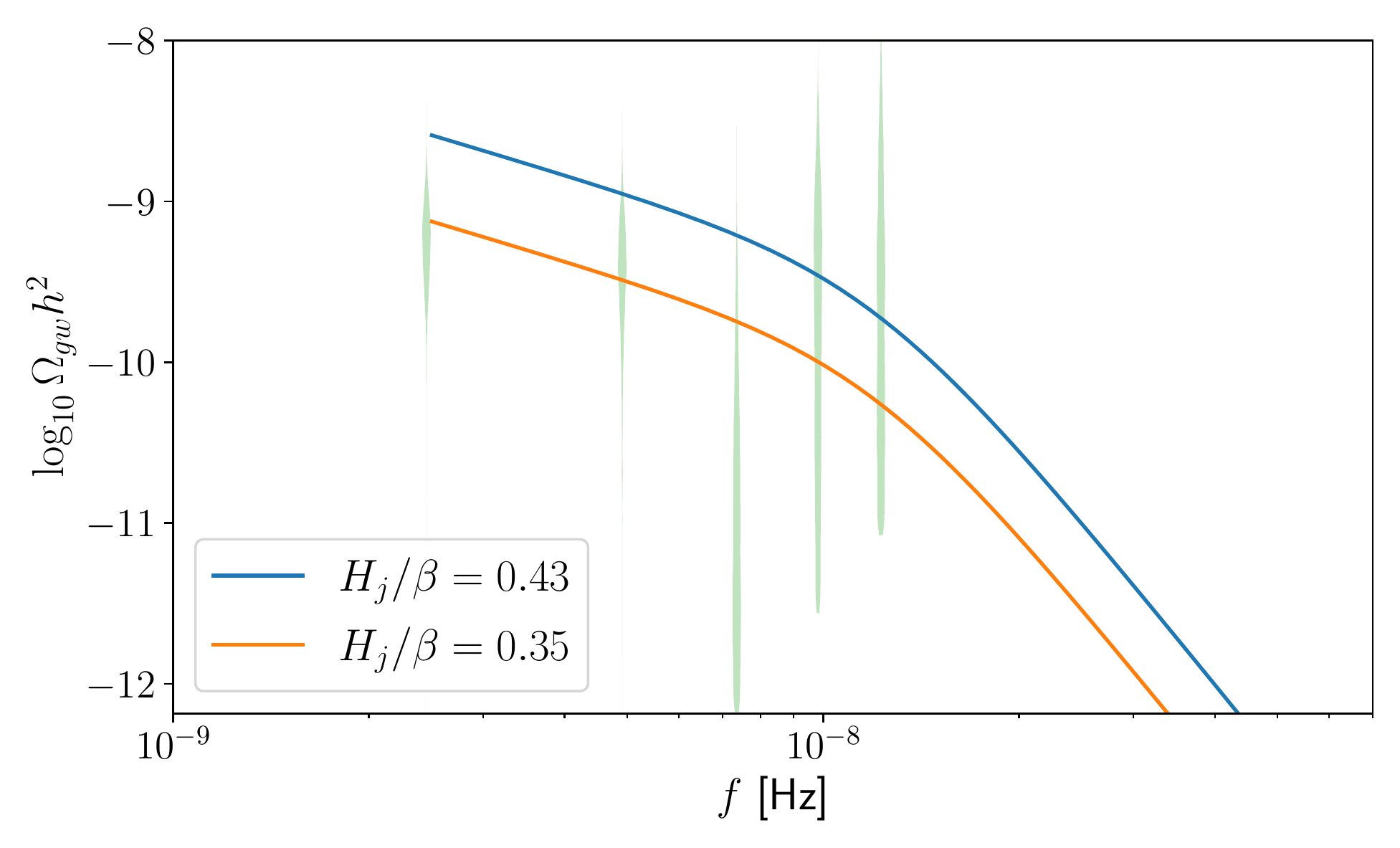}
\includegraphics[width=0.45\textwidth]{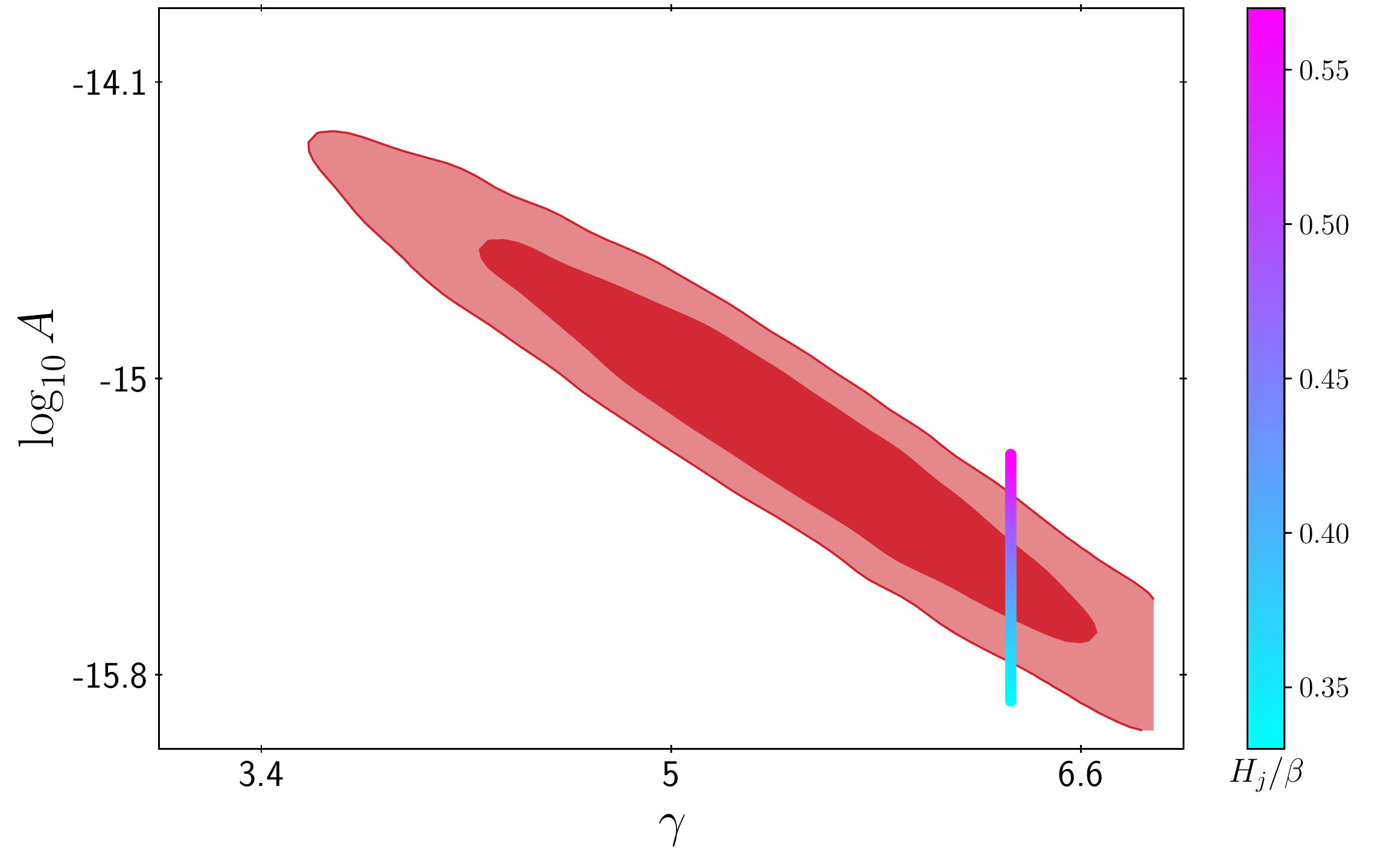}
\caption{Left panel: The stochastic GWB (\ref{GW0}) at the
NANOGrav frequency band. Violin points plot marginalized posteriors of the equivalent amplitude of the first five frequency bins in a free-spectrum analysis of the NANOGrav 12.5-yr data \cite{Arzoumanian:2020vkk}. We set ${H_{j}}/{\beta}=0.35,\, 0.43$,
respectively, for $f_{c}\simeq 1.2\times 10^{-8}$Hz. There is a
low-frequency cutoff $f_{cutoff}\simeq 2.5\times 10^{-9}$Hz, since
the initial GWB (\ref{Omega-t}) yielded by PT must be sub-horizon.
Right panel: the $1\sigma$ and $2\sigma$ contours of the amplitude
$A$ and tilt $\gamma$ for a power-law NANOGrav signal. The
colorbar represents the values of $\{$$A$-$\gamma$$\}$ for
(\ref{GW0}) with respect to $H_j/\beta$. } \label{fig-Omega}
\end{figure}

It is interesting to connect (\ref{pt}) with recent experimental results. The analysis result of NANOGrav 12.5-yr
data is modeled as a signal with the characteristic strain
amplitude \cite{Arzoumanian:2020vkk} $h_c(f)=A\lf({f/
f_{yr}}\rt)^{(3-\gamma)/2}$. Thus the corresponding energy
spectrum $\Omega_{GW}={2\pi^2\over 3H_0^2}f^2h_c^2(f)$ is \be
\Omega_{GW}={2\pi^2\over 3H_0^2}f_{yr}^2A^2 \lf({f\over
f_{yr}}\rt)^{5-\gamma}, \label{GW1}\ee where $f_{yr}=1/yr$.
Present energy spectrum $\Omega_{GW}(\tau_{0})$ of GWs (\ref{pt})
is \cite{Turner:1993vb}
\begin{equation}
\Omega_{GW}(\tau_{0})=\frac{k^{2}}{12
a_0^2H^2_0}P_{T}(k)\lf[\frac{3
\Omega_{{m}}j_1(k\tau_0)}{k\tau_{0}}\sqrt{1.0+1.36\frac{k}{k_{\text{eq}}}
+2.50\left( \frac{k}{k_{\text{eq}}}\right) ^{2}}\rt]^2,
\label{GW0}\end{equation} see also
\cite{Boyle:2005se,Zhao:2006mm,Kuroyanagi:2014nba}. Here,
$\Omega_m=\rho_m/\rho_c$ and $\rho_{{c}}=3H^{2}_0/\big(8\pi
G\big)$ is the critical energy density, $1/k_{eq}$ is the Hubble
scale at matter-radiation equality. According to (\ref{GW1}), if
the NANOGrav signal is regarded as the stochastic GWB (\ref{GW0}),
we have \be A=\lf({3H_0^2\Omega_{GW}(f_0)\over
2\pi^2f_{yr}^2}\rt)^{1/2}\lf({f_{yr}\over
f_0}\rt)^{(5-\gamma)/2}\quad for \quad {f_0= 5\times 10^{-9} Hz}
\label{A}\ee and $\gamma\simeq 6$. We plot
$\Omega_{GW}(\tau_{0})$ in the left panel of Fig-\ref{fig-Omega}
for ${\Lambda_{j+1}/ \Lambda_{j}=0.2}$. Thus if
${H_{j}/\beta}\simeq 0.4$, we have $\Omega_{GW}\sim 10^{-9}$ with
the frequency at $(f_{cutoff},1.4f_{cutoff}\beta/H_j)$, which
might explain the NANOGrav result for
$f_{cutoff}\simeq 2.5\times 10^{-9}Hz$.

We plot the $1\sigma$ and $2\sigma$ NANOGrav contours of
$\{$$A$-$\gamma$$\}$ in the right panel of
Fig-\ref{fig-Omega}. As expected, if ${H_{j}/\beta}\sim 0.4$, the
short-lived dS decay predicts $A\sim 10^{-15}$ and $\gamma\simeq
6$, which fits the NANOGrav data at $1\sigma$ level. The
subsequent PT will occur after the $(j+1)$-th inflation with lower
dS energy $\Lambda_{j+1}$ lasting for efolds $N_{j+1}<\ln{\left(
M_P\over H_{j+1}\right) }$, see Fig-\ref{fig-PT}, so a ``reddened"
stochastic GWB will be also possibly imprinted in high frequency,
which might be detectable by GW detectors such as LISA, TAIJI and TianQin. It
is noted that the stochastic GWB in the cosmic string scenario
corresponds to $\gamma \lesssim 5$
\cite{Ellis:2020ena,Blasi:2020mfx}, see also
\cite{Vaskonen:2020lbd,DeLuca:2020agl,Nakai:2020oit,Buchmuller:2020lbh,Addazi:2020zcj,Kohri:2020qqd,Ratzinger:2020koh,Samanta:2020cdk,Bian:2020bps,Neronov:2020qrl}
for other GW sources, while ours is $\gamma\simeq 6$.

In summary, we showed that the NANOGrav signal might be telling us
the existence of short-lived dS vacua. As suggested by the
swampland conjectures
\cite{Ooguri:2006in,Obied:2018sgi,Bedroya:2020rac}, the dS space
is highly unstable. During inflation, assuming the swampland conjectures, dS cascade seems to be
inevitably present. We found that the GWs yielded by the
corresponding PT will be reddened by subsequent dS expansion,
which results in a stochastic GWB at low frequency, compatible
with the recent NANOGrav result at $1\sigma$ level. It should be
mentioned that the NANOGrav collaboration did not claim a
detection of GWs, since the signal seems not exhibiting quadrupole
correlations. However, our spectrum (except the amplitude) of the
stochastic GWB is universal for the dS decay during inflation.
Though the model we consider is quite simplified, it highlights an
unexpected point that the short-lived dS vacua, emerging in a
consistent UV-complete theory, might imprint unique voiceprint in our observable Universe.

\textbf{Acknowledgments}

Hao-Hao Li thanks Yu-Tong Wang for helpful discussions. YSP is
supported by NSFC, Nos.12075246, 11690021. The contour plot in
Fig-\ref{fig-Omega} is produced using the publicly available
code\footnote{\url{https://github.com/nanograv}} Enterprise,
Enterprise-extension together with PTMCMC
\cite{justin_ellis_2017_1037579},
libstempo\footnote{\url{http://vallis.github.io/libstempo/}} and
TEMPO2\footnote{\url{https://bitbucket.org/psrsoft/tempo2/src/master/}}.

\appendix
\section*{Appendix}

In this Appendix, we present a model of multi-stage
inflation in Fig.\ref{fig-PT}, which is compatible with the swampland
conjectures (SCs). Then we estimate the parameters $ \beta $ and
$H_j/\beta $.

\begin{figure}
    \centering
    \includegraphics[scale=2,width=0.8\textwidth]{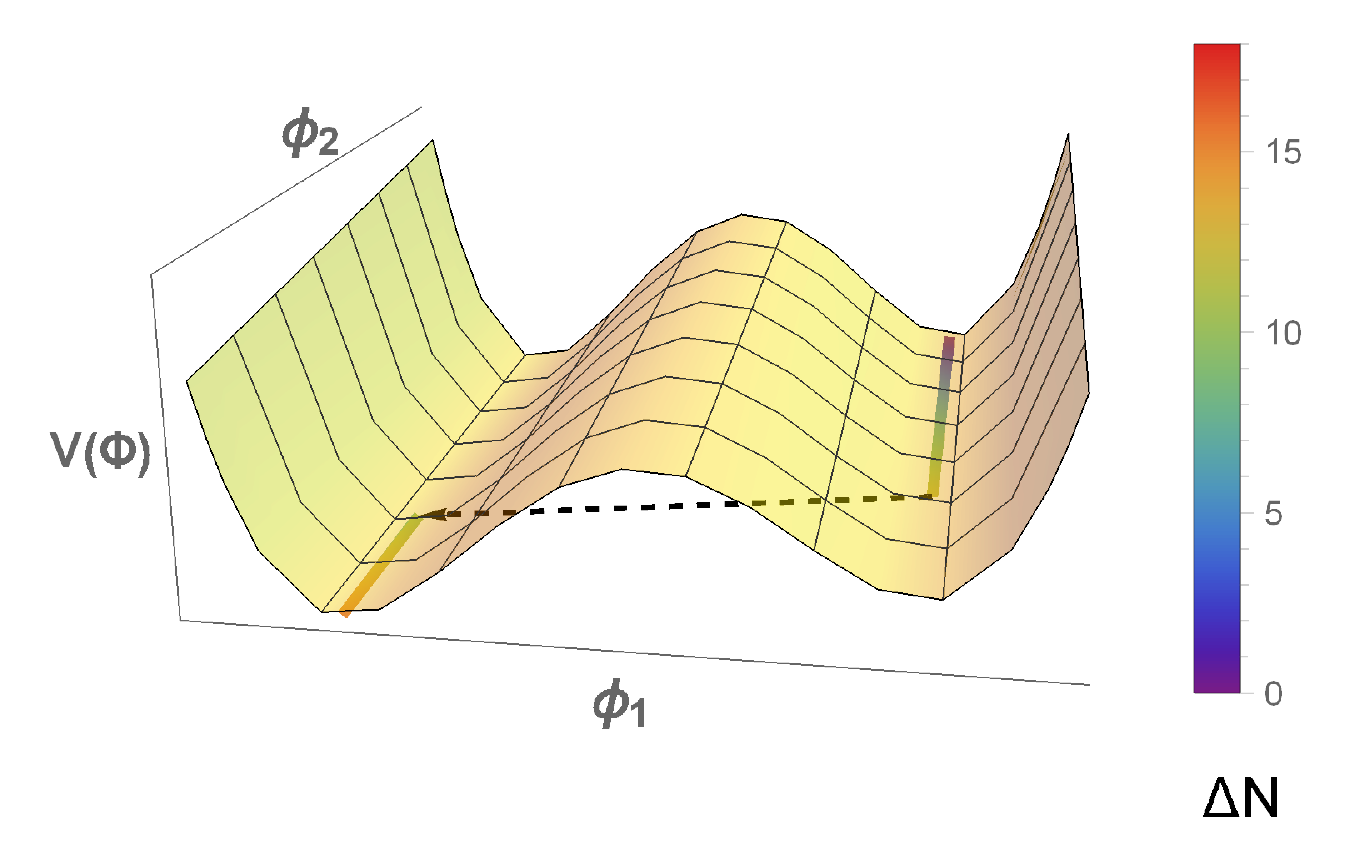}
    \caption{Illustration of $\Phi$ evolution in the field space with color coding for the e-folding number.
    }
    \label{bivacua}
\end{figure}

We consider a set of canonical scalar fields $\Phi=(\phi_1,\phi_2)$ with the potential as follows
\begin{gather}
    V(\Phi)= V_{PT}(\phi_1,\phi_2) + V_{inf}(\phi_2) ,\tag{A.1}
\end{gather}
and
\begin{gather}
    V_{PT}(\phi_1,\phi_2) =\left[  A(\phi_2)\left( \frac{\phi_1}{\phi_*}\right)^4 + B(\phi_2)\left( \frac{\phi_1}{\phi_*}\right)^3+ C(\phi_2)\left( \frac{\phi_1}{\phi_*}\right)^2 \right]\Lambda_j  + \Lambda_{j+1},  \tag{A.2}
\end{gather}
with coefficients $A(\phi_2) = (-8+16\sigma)$, $B(\phi_2) = (14-32\sigma)$, $ C(\phi_2) = (-5+16\sigma)$ and $ \sigma = \lambda \left(\alpha - \ln(\phi_2/m_2) \right)$. Here, $m_2$ has the mass dimension. The parameters  $\phi_* , \lambda , \alpha $ are constant, $ \phi_1 $ is the field responsible for PT, $ \phi_2 $ is the inflation field, $ \sigma $ depends on $ \phi_2 $ and is constant in the $ \phi_1 $-direction.  $V_{PT}$ has two minima along the $ \phi_1 $-direction with $ \Lambda_{j}+\Lambda_{j+1} $ and $ \Lambda_{j+1} $ at $ \phi_1 = \phi_* $  and $ \phi_1 = 0 $ respectively. The height of the barrier between two minima is about $ \sigma \Lambda_{j} $. $V_{inf}$ drives inflation on $\phi_2$-direction at $ \phi_1 = \phi_* $ . Before PT, the universe is dominated by $ V_{PT} \simeq \Lambda_{j} $ which requires $ V_{inf} < V_{PT} $ and $ \Lambda_{j+1} \ll \Lambda_{j} $. We plot the potential in Fig-\ref{bivacua}.

In the following, we show this model is established in a stable dS space that can satisfy the SCs, especially the de Sitter swampland conjecture (dSC). The refined dSC in mutli-field requires either \cite{Ooguri:2018wrx}
\begin{gather}
M_P \frac{\lvert \nabla V \rvert}{V} = M_P\frac{\sqrt{G^{ab}\partial_a V\partial_b V}}{V} > c \qquad \text{or} \qquad M_P^2\frac{\min (\nabla_a \nabla_b V)}{V} < -c', \label{A3} \tag{A.3}
\end{gather}
for some constants $ c,c' \sim {\cal O}(1) $.  For canonical fields, the field space metric $G_{ab}=\delta_{ab}$. In our model, when inflation begins on the $ \phi_2 $-direction, $ \frac{\partial V_{PT}}{\partial_{\phi_1}} \Big|_{\phi_1 = \phi_*} = 0 $ and $ V_{PT}(\phi_1 = \phi_*) \simeq \Lambda_{j} $, where $ \Lambda_{j+1} $ is much smaller that can be ignored. So Eq. \eqref{A3} becomes
\begin{gather}
M_P \frac{\lvert \nabla V \rvert}{V} \simeq \frac{M_P  V_{inf}^\prime(\phi_2)}{\Lambda_{j} + V_{inf}(\phi_2)}>c\qquad \text{or} \qquad M_P^2\frac{\min (\nabla_a \nabla_b V)}{V}\simeq\frac{M_P^2V''_{inf}(\phi_2)}{\Lambda_{j} + V_{inf}(\phi_2)}<-c',\tag{A.4} \label{A4}
\end{gather}
where the prime denotes derivation with respect to $ \phi_2 $. Being compatible with the Eq.\eqref{A4} requires fast-roll inflation on the $\phi_2$-direction.

If we assume the $j$-th stage inflation begins near the GUT energy scale $H_j \sim 10^{15}\text{GeV} $, then the e-folding number in fast-roll inflation can be as large as $\Delta N\simeq10$ \cite{Kallosh:2001gr}. The e-folding number is bounded by TCC to be $ \Delta N_j^{TCC} = \ln\left( \frac{M_P}{H_j} \right) \sim 10$, which is compatible with our model. Thus several stages of inflation separated by PT are enough to give a total e-folding number as expected while also being compatible with SCs.

    The nucleating rate of bubble is $ \Gamma \sim e^{-S_E} $, and in the thin-wall approximation\cite{Coleman:1977py} we have
    \begin{gather}
    S_E = \frac{27 \pi^2 S_1^4}{2 (\Delta \Lambda)^3}, \label{A5} \tag{A.5}
    \end{gather}
    with the bubble wall surface tension $ S_1 \simeq\sqrt{2\sigma\Lambda_{j}}\phi_* $, thus ($\Lambda_{j}\gg\Lambda_{j+1}$)
    \begin{equation}
    \Gamma \propto \exp\left[-54\pi^2\sigma^2\frac{\phi_*^4}{\Lambda_{j}}\right]. \label{A6} \tag{A.6}
    \end{equation}

 The equation of motion of $ \phi_2 $ is
\begin{equation}
 \ddot{\phi_2}+ 3H_j \dot{\phi_2} + V_{inf}^\prime = 0.  \tag{A.7}
\end{equation}
In the fast roll scenario, $ \ddot{\phi_2} \gg 3H_j\dot{\phi_2} $ so the Hubble friction term can be ignored. For a fast-rolling regime with $ \lvert V_{inf}^{''} \rvert = 4H_j^2 $, one has
\begin{equation}
    \phi_2\simeq \phi_2^i e^{H_j(t-t^i)} ,  \tag{A.8}
\end{equation}
where $ \phi_2^i $ is the initial value of $ \phi_2 $. With a proper choice of $\phi_2^i$, one can have $\Delta\Phi=\Delta\phi_2<M_p$ within one stage of inflation. Taking $m_2=\phi_2^i$ for convenience, one has $ \sigma^2 = \lambda^2 \alpha^2 \left(1 - \frac{2}{\alpha} H_j(t-t^i) \right)   $ with $ \alpha \gg H_j(t-t^i) $. Inserting into Eq. \eqref{A6}, we get $\Gamma\propto \exp\left[\beta(t-t^i) \right]$ and
\begin{equation}
    \frac{H_j}{\beta}\simeq \frac{\Lambda_{j}}{108\pi^2 \alpha \lambda^2  \phi_*^4 }.  \tag{A.9}
\end{equation}
Thus by a suitable choice of parameters $\phi_*$,
$\alpha$ and $ \lambda $, which affects the width and height of the
potential barrier, it is possible to have some values of
$H_j/\beta$ compatible with the NANOGrav 12.5-yr observation plotted in Fig-\ref{fig-Omega}.

 \end{document}